\documentclass[pra,twocolumn,superscriptaddress,notitlepage,showpacs,showkeys]{revtex4-1}
\usepackage{graphicx,amsmath,amsfonts,amssymb,upgreek,txfonts,color}
\usepackage[colorlinks,linkcolor=blue,citecolor=blue,urlcolor=blue,breaklinks=true]{hyperref}
\usepackage[utf8x]{inputenc} 

\begin{document}

\title{Investigation of intrinsic nonlinear effects in driven-dissipative optomechanical systems using the generalized linear response theory}

\author{B. Askari} 
\address{Laser and Plasma Research Institute, Shahid Beheshti University, Tehran 19839-69411, Iran}

\author{A. Dalafi} 
\email{a\_dalafi@sbu.ac.ir}
\address{Laser and Plasma Research Institute, Shahid Beheshti University, Tehran 19839-69411, Iran}

\author{M. J. Kazemi} 
\address{Laser and Plasma Research Institute, Shahid Beheshti University, Tehran 19839-69411, Iran}

\begin{abstract}
In this article, we study the effects of intrinsic nonlinear optomechanical interaction on the linear response of a driven-dissipative optomechanical system to a weak time-dependent perturbation. By calculating the linear response of the cavity optical mode to a weak probe laser in the framework of the generalized linear response theory, it is shown how the Stokes and anti-Stokes sideband amplitudes as well as the power reflection coefficient, and the density of states of the cavity optical mode are expressed in terms of photonic retarded Green's functions. Then, we derive the equations of motion of retarded Green's functions of the system from nonlinear quantum Langevin equations and solve them. It is shown that for a single-photon optomechanical coupling of the order of the cavity linewidth, the nonlinear effect does not manifest itself unless the system satisfies a resonance condition, where the frequency of the upper normal mode of the system is twice that of the lower one. Based on the generality of the present approach which works at all regimes, the validity of linearization approximation is also confirmed at the off-resonance regime. 

\end{abstract}


\maketitle
\section{Introduction}
Recent theoretical and experimental developments in quantum optomechanics \cite{Aspelmeyer,Bowen book} have provided suitable conditions to observe the nonlinear nature of radiation pressure interaction between the optical and mechanical modes. Although the optomechanical interaction is nonlinear intrinsically, most achievements like ground-state cooling of the mechanical vibrations \cite{Teufel}, displacement and force sensing \cite{Tsa X 2012,Wimm 2014,aliNJP,Fani2020,aliDCEBECforce,aliAVSbook2020}, and entanglement generation \cite{Palom} have been obtained in strongly driven regime where the dynamics is effectively linear. Nevertheless, there are interesting features like generation of nonclassical states and several others which occur in the nonlinear regime \cite{Rabl,Signature,singlephoton}.

The effects of nonlinear optomechanical interaction do not arise in driven-dissipative optomechanical systems (OMSs) unless the single-photon optomechanical coupling is of the order of the cavity linewidth. However, achieving such a large single-photon optomechanical coupling has been very challenging experimentally with few exceptions like superconducting circuits \cite{SupCon}, silicon-based optomechanical crystals \cite{OMC}, and hybrid optomechanical systems containing cold atoms or Bose-Einstein condensate (BEC) \cite{Kanamoto,Brenn Science,Ritter Appl. Phys. B}. In such hybrid OMSs the fluctuation of the collective excitation of the atomic field behaves like an effective mechanical mode  which is coupled with the radiation pressure of cavity optical field \cite{Dalafi Dispersive,Dalafi CK,Dalafi 2BEC,Dalafi EQC}.

Interestingly, many of the essential intrinsic properties and dynamical behaviors of such driven-dissipative quantum systems can be straightforwardly determined by the study of their linear response to a weak time-dependent perturbation which can be formulated either in the Schr{\"o}dinger picture through the master equation approach \cite{greenScarlatella1,banPRA17,shenPRA17} or in the Heisenberg picture through the quantum Langevin equations (QLEs) \cite{shenOptLett18}. Furthermore, there are linear response approaches based on Green's function equations of motion \cite{hashem,Tong} which have been mostly formulated for closed quantum systems in contact with a thermal reservoir at finite temperature. Nevertheless, a similar approach based on Green's function equation of motion has been recently introduced in the framework of theory of open quantum system, named the generalized linear response theory (GLRT) \cite{aliGreen}.

Based on the GLRT, one can obtain  a set of equations of motion for the open quantum system Green's functions which are derived in the Heisenberg picture through the QLEs. Furthermore, the GLRT clarifies the relations between the system susceptibilities and Green’s functions. In this way, one can describe many interesting quantum optical phenomena like normal mode splitting \cite{Dobr,Grob,NMS Huang}, electromagnetically induced transparency (EIT) \cite{Imamoglu}, slow light realization \cite{Harris, mik1}, optomechanically induced transparency (OMIT) \cite{OMIT1,OMIT2,oMITReview2018,oMIT Ann Phys}, optomechanically induced gain (OMIG) \cite{OMIG1,mik2}, and Fano resonance \cite{Fano rev,Fano Qu,Fano What,Fano Abbas} in terms of the open system Green's functions. These phenomena have been recently studied \cite{askari, Negat Ali} in bare and hybrid OMSs in the framework of GLRT in the linear regime, where the cavity photon number is large and the single-photon optomechanical coupling is small compared with the cavity linewidth and mechanical frequency.

In the present work, we study the linear response of a bare OMS in the framework of GLRT beyond the linearization approximation where the intrinsic nonlinear optomechanical interaction is no longer ignorable. For this purpose, we derive the equations of motion of the system Green's functions from nonlinear QLEs in the framework of the theory of open quantum systems. In spite of previous papers \cite{aliGreen,askari, Negat Ali}, where the system has been studied in the linearization approximation, the great challenge of the present work in applying the GLRT for a nonlinear OMS is the manifestation of an infinite hierarchy in the Green's functions equations of motion. Using Wick's theorem \cite{Louisell book} and considering the 3-fold correlations as new variables, we demonstrate that the hierarchy can be truncated by neglecting 4-fold correlations, resulting in a system of linear differential equations with a dimensionality higher than that of the nonlinear QLEs for the Green’s functions.

Our results show that in the nonlinear regime, where the single-photon optomechanical coupling is of the order of cavity linewidth, the intrinsic nonlinear effect of optomechanical interaction manifests itself as a hybridization of the second peak of the optical density of states only when the system satisfies a resonance condition in which the frequency of the upper normal mode of the system is twice that of the lower one. Interestingly, in the off-resonance regime, the nonlinear effect does not manifest even when the single-photon optomechanical coupling becomes of the order of cavity linewidth. In this way, the validity of the linearization approximation in quantum optomechanics is confirmed at the off-resonance regime.

It should be reminded that there is also another method based on Keldysh Green's function approach to study the nonlinear interaction effects in strongly driven OMSs \cite{Lemonde2013,Lemonde2015} in which the nonlinear photonic Green's function is obtained from the polariton retarded ones through a perturbative calculation of the Keldysh self-energy corresponding to the nonlinear interaction. In spite of its rigor and generality, the Keldysh Green's function approach becomes very complicated when combined with the theory of open quantum systems because not only the open quantum system should be described in terms of polariton basis but also the modeling of environment should be reformulated. Besides, a basis transformation is required to extract the photonic Green's function from the polariton ones at the end of calculation. The great advantage of the GLRT method is that all the calculations are done in the original basis so that the photonic (and phononic) Green's functions are derived directly from their corresponding equations of motion without any necessity of basis transformation and reformulating the environment modeling.

The paper is structured as follows: In Sec. \ref{Hamiltonian}, the open quantum system Hamiltonian and its dynamics is described using the theory of open quantum systems. In Sec. \ref{GLRT}, the linear response of the nonlinear OMS is described in the framework of the GLRT, and in Sec.\ref{EOM} the equations of motion of retarded Green's functions are derived from the nonlinear QLEs. The obtained results based on the GLRT are discussed in Sec.\ref{RD}, and finally our conclusions are summarized in Sec. \ref{cln}.

\section{Open system Hamiltonian and dynamics}\label{Hamiltonian}
The system under consideration is a bare OMS consisting of a Fabry-Perot cavity with length $L$, damping rate $\kappa$, and resonance frequency $ \omega_{0} $, whose right mirror is a mechanical oscillator with the natural frequency $\omega_m$, and damping rate $\gamma$. The cavity is pumped by a strong coupling laser with frequency $ \omega_{c} $ at the rate of $\eta$ through the left (fixed) mirror, which is responsible for the generation of optomechanical coupling between the cavity optical mode and the mechanical oscillator. Such an OMS is described by the following Hamiltonian in the frame rotating at the coupling laser frequency
\begin{equation}\label{HOMS}
H_{OMS}=\hbar\Delta_c a^{\dagger} a +\hbar\omega_m b^{\dagger}b+\hbar g_0 a^{\dagger}a (b+b^{\dagger}) +\hbar (\eta^{\ast} a + \eta a^{\dagger}),\nonumber\\
\end{equation}
where $\Delta_c=\omega_{0}-\omega_{c}$ is the detuning between the coupling laser frequency and cavity resonance. The first two terms are, respectively, the free energies of the optical and mechanical modes, and the third term is the radiation pressure (optomechanical) interaction with the single-photon optomechanical coupling $g_0$. The last term is the optical drive of the cavity by the coupling laser.

The dynamics of the system is described by the following set of nonlinear QLEs in the framework of theory of open quantum systems\cite{Bowen book} 
\begin{subequations}
	\begin{eqnarray}
	\dot{a}&=&-(i\Delta_c+\kappa/2) a-i g_0 a (b + b^{\dagger})-i\eta+\sqrt{\kappa}\delta a_{in},\label{NLQLEa}\\
	\dot{b}&=&-(i\omega_{m}+\gamma/2) b -ig_0 a^{\dagger}a+\sqrt{\gamma}\delta b_{in}.\label{NLQLEb}
	\end{eqnarray}
\end{subequations}
As is seen from Eqs.(\ref{NLQLEa})-(\ref{NLQLEb}), the system dynamics is affected by two uncorrelated quantum noise sources due to the interaction of the system with the optical and mechanical reservoirs. The optical input vacuum noise $\delta a_{in}$ arising from all the optical modes outside the cavity satisfies the Markovian correlation functions, i.e., $\langle\delta a_{in}(t)\delta a_{in}^{\dagger}(t^{\prime})\rangle=(n_{ph}+1)\delta(t-t^{\prime})$, $\langle\delta a_{in}^{\dagger}(t)\delta a_{in}(t^{\prime})\rangle=n_{ph}\delta(t-t^{\prime})$ with the average thermal photon number $n_{ph}$ which is nearly zero at optical frequencies \cite{Zubairy,Gardiner}. The mechanical quantum noise operator $\delta b_{in}$ arising from the mechanical reservoir satisfies similar correlation function relations as $\langle\delta b_{in}^{\dagger}(t)\delta b_{in}(t^{\prime})\rangle=n_{th}\delta(t-t^{\prime})$ with the average thermal phonon number $n_{th}=[\exp(\hbar\omega_{m}/k_B T)-1]^{-1}$.

Since every quantum field operator can be written as a classical mean-field plus a quantum fluctuation operator, one can write $a=\alpha+\delta a$ for the optical field and $b=\beta+\delta b$ for the mechanical mode. By substituting these relations for the optical and mechanical fields in Eqs.(\ref{NLQLEa})-(\ref{NLQLEb}), one can obtain the following set of equations of motion for quantum field fluctuations
\begin{subequations}
	\begin{eqnarray}
	\delta\dot{a}&=&-(i\Delta+\kappa/2)\delta a-i(g+g_0\delta a) (\delta b+\delta b^{\dagger})+\sqrt{\kappa}\delta a_{in},\label{Flca}\\
	\delta\dot{b}&=&-(i\omega_{m}+\gamma/2)\delta b -ig(\delta a+\delta a^{\dagger})-ig_0\delta a^{\dagger}\delta a+\sqrt{\gamma}\delta b_{in},\label{Flcb}\nonumber\\
	\end{eqnarray}
\end{subequations}
where $g=g_0\alpha$ is the effective optomechanical coupling. It should be noted that if the phase of coupling laser (phase of the complex variable $\eta$) is chosen as $\theta = tg^{-1}
(-\kappa/2\Delta)$  then the steady state value of the optical mean field $\alpha=-\eta/(\Delta-i\kappa/2)$ becomes a real value. Furthermore, $\Delta=\omega_0-\omega_c + 2g_0\beta_R$ is the effective cavity detuning, where $\beta_{R}$ is the real part of the mechanical mean field whose steady state value is given by $\beta=-g_0|\alpha|^2/(\omega_m-i\gamma/2)$. As is seen from Eqs.(\ref{Flca})-(\ref{Flcb}), the terms with the coefficient $g_0$ are nonlinear. Obviously, in the limit of $g_0\ll g$, and $g_0\ll\kappa$ the linearized form of QLEs are derived from Eqs.(\ref{Flca})-(\ref{Flcb}).

Furthermore, in the framework of the theory of open quantum systems it can be easily shown that the nonlinear equations of motion of quantum fluctuations, i.e., Eqs.(\ref{Flca})-(\ref{Flcb}), are derived from the following effective Hamiltonian
\begin{eqnarray}\label{HS}
H_S &=& \hbar\Delta \delta a^{\dagger}\delta a+\hbar\omega_m\delta b^{\dagger}\delta b+ \hbar g (\delta a+\delta a^{\dagger})(\delta b+\delta b^{\dagger})\nonumber\\
&&+\hbar g_0 \delta a^{\dagger}\delta a (\delta b+\delta b^{\dagger}),
\end{eqnarray}
which can be considered as the open system Hamiltonian. As is seen, the last term of Eq.(\ref{HS}) is the nonlinear interaction which leads to the manifestation of the nonlinear terms in the equations of motion of quantum fluctuations, i.e., the terms with coefficient $g_0$ in  Eqs.(\ref{Flca})-(\ref{Flcb}). It should be noted that this nonlinearity is ignorable in the (linear) regime where $\alpha\gg 1$ and $g_0\ll\kappa,\omega_m$; otherwise it should be taken into account because of its important outcomes.

Now, by introducing the vector of quantum fluctuations as $\boldsymbol{u}^T(t)=[\delta a(t),\delta a^{\dagger}(t),\delta b(t),\delta b^{\dagger}(t)]$, the nonlinear equations of motion (Eqs.(\ref{Flca})-(\ref{Flcb})) can be written in the following compact form
\begin{eqnarray}\label{compEOM}
\dot u^i(t)=\chi^{i}_{m} u^m (t) + \Gamma^{i}_{mn} u^{m}(t)u^{n}(t)+\xi^i(t),
\end{eqnarray}
where $\boldsymbol{\xi}^T(t)=[\sqrt{\kappa}\delta a_{in}(t),\sqrt{\kappa}\delta a^{\dagger}_{in}(t),\sqrt{\gamma}\delta b_{in}(t),\sqrt{\gamma}\delta b^{\dagger}_{in}(t)]$ is the vector of quantum noises, and $\boldsymbol{\chi}$ is the coefficient matrix corresponding to the linear part of equations of motion which is given by

\begin{equation}\label{chi}
\boldsymbol{\chi}=\left(\begin{array}{cccc}
-\frac{\kappa}{2}-i\Delta & 0 & -ig & -ig \\
0 & -\frac{\kappa}{2}+i\Delta & ig & ig \\
-ig & -ig & -\frac{\gamma}{2}-i\omega_{m} & 0 \\
ig & ig & 0 & -\frac{\gamma}{2}+i\omega_{m}\\
\end{array}\right),
\end{equation}
while $\boldsymbol{\Gamma}^{i}$'s, with $i=1,...,4$ are the coefficient matrices corresponding to the nonlinear parts of the equations of motion which are given by

\begin{align}
	&\boldsymbol{\Gamma}^1=\left(\begin{array}{cccc}
	0 & 0 & -ig_0 & -ig_0 \\
	0 & 0 & 0 & 0 \\
	0 & 0 & 0 & 0 \\
	0 & 0 & 0 & 0\\
	\end{array}\right),
	&\boldsymbol{\Gamma}^2=\left(\begin{array}{cccc}
	0 & 0 & 0 & 0 \\
	0 & 0 & ig_0 & ig_0 \\
	0 & 0 & 0 & 0 \\
	0 & 0 & 0 & 0\\
	\end{array}\right),\nonumber
	\\
	&\boldsymbol{\Gamma}^3=\left(\begin{array}{cccc}
	0 & 0 & 0 & 0 \\
	-ig_0 & 0 & 0 & 0 \\
	0 & 0 & 0 & 0 \\
	0 & 0 & 0 & 0\\
	\end{array}\right),
	&\boldsymbol{\Gamma}^4=\left(\begin{array}{cccc}
	0 & 0 & 0 & 0 \\
	ig_0 & 0 & 0 & 0 \\
	0 & 0 & 0 & 0 \\
	0 & 0 & 0 & 0\\
	\end{array}\right).\nonumber
	\\
\end{align}
It should be noted that in Eq.(\ref{compEOM}), we have used the Einstein's summation convention.

\section{Linear response of the nonlinear OMS in the framework of the GLRT}\label{GLRT}
The question we are going to address is how an OMS with the nonlinear Hamiltonian of Eq.(\ref{HS}) whose dynamics is fully described by the nonlinear equations of motion (Eqs.(\ref{Flca})-(\ref{Flcb})), responds to an optical time-dependent perturbation especially in the regime where the nonlinear interaction is not ignorable. For this purpose, we assume that the OMS is also driven by a weak probe laser through the fixed mirror at the rate $\zeta$ whose absolute value is much smaller than that of the coupling laser. The probe laser generates a weak time-dependent perturbation which is given by the following potential in the frame rotating at the coupling laser frequency
\begin{equation}\label{Vt}
V(t)=\hbar\zeta\delta a e^{i\omega_{pc}t}+\hbar\zeta^\ast\delta a^{\dagger} e^{-i\omega_{pc} t}
\end{equation}
where $\omega_{pc}=\omega_p-\omega_c$  is the detuning between the probe and coupling lasers frequencies. 

In the framework of the theory of open quantum systems, the Hamiltonian of the total system (the system plus the reservoir) is given by
\begin{subequations}
	\begin{eqnarray}
	H &=& H_0 + V(t),\\
	H_0 &=& H_S + H_\kappa + H_\gamma,
	\end{eqnarray}
\end{subequations}
where $H_\kappa$ $(H_\gamma)$ is the Hamiltonian of the optical (mechanical) reservoir together with its interaction with the system optical (mechanical) mode. As previously stated $H_S$ and $V(t)$ are, respectively, the Hamiltonian of the open system (Eq.(\ref{HS})), and the time-dependent perturbation generated by the probe laser (Eq.(\ref{Vt})). 

Now, using the GLRT \cite{aliGreen}, we can obtain the linear response of the present OMS (having intrinsic nonlinearity) to a weak time-dependent perturbation. In this way, the response of the optical field fluctuation of the OMS to the external time-dependent perturbation can be obtained as follows 
\begin{eqnarray}\label{Mat}
	\langle\delta a(t)\rangle=\langle\delta a\rangle_0+\zeta\int_{-\infty}^{+\infty} dt^{\prime} G_R^{aa}(t-t^{\prime}) &&e^{i\omega_{pc}t^{\prime}}\nonumber\\
	+\zeta^\ast\int_{-\infty}^{+\infty} dt^{\prime} G_R^{aa^\dag}(t-t^{\prime}) e^{-i\omega_{pc}t^{\prime}},\label{mat}
\end{eqnarray}
where $\langle\delta a (t)\rangle$ is the expectation value of the optical field fluctuation in the Heisenberg picture in the presence of the time-dependent perturbation while $\langle\delta a\rangle_0=0$ is the steady state mean-value of the optical field fluctuation in the interaction picture (which is equivalent to the Heisenberg picture in the absence of the time dependent perturbation). Besides, the optical field retarded Green's functions are defined as \cite{Ryndyk,Coleman,Flensberg,Stefanucci,Economou}
\begin{subequations}
	\begin{eqnarray}
	G_R^{aa}(t-t')&=&-i\theta(t-t')\langle [\delta a(t),\delta a(t')]\rangle_0,\label{Gaa}\\
	G_R^{aa^\dag}(t-t')&=&-i\theta(t-t')\langle [\delta a(t),\delta a^\dag(t')]\rangle_0\label{Gaad},
	\end{eqnarray}
\end{subequations}
where $\theta(t-t')$ is the Heaviside (step) function.
 It should be reminded that the subscript 0 indicates that all the expectation values should be calculated in the steady-state of the system in the absence of the perturbation as was mentioned before. Therefore, the time evolution of $\delta a(t)$ in Eqs.(\ref{Gaa})-(\ref{Gaad}) is obtained from the nonlinear QLEs given by Eq.(\ref{compEOM}). On the other hand, using the definition of the Fourier transform of the Green's function, i.e., $\tilde{G}(\omega)=\int_{-\infty}^{+\infty}d\tau G(\tau) e^{i\omega\tau}$, Eq.(\ref{mat}) can be rewritten as 
\begin{eqnarray}
		\langle\delta a(t)\rangle&=&\zeta^\ast \tilde G_R^{aa^\dag}(\omega_{pc}) e^{-i\omega_{pc}t}+\zeta \tilde G_R^{aa}(-\omega_{pc}) e^{i\omega_{pc}t}.
\end{eqnarray}

Since $\langle a(t)\rangle=\alpha+\langle\delta a(t)\rangle$ is the expectation value of the optical field in the rotating frame, the response of the optical field to the time-dependent perturbation is obtained as
\begin{eqnarray}\label{matL}
	\langle a(t)\rangle = \alpha +\zeta^\ast \tilde G_R^{aa^\dag}(\omega_{pc}) e^{-i\omega_{pc}t} + \zeta\tilde G_R^{aa}(-\omega_{pc}) e^{i\omega_{pc}t},\label{Ra}
	\end{eqnarray}
As is seen from Eq.(\ref{Ra}), the optical mode has a central band (the first term) and two sidebands, the so-called anti-Stokes (the second term) and Stokes (the third term) sidebands. 

Based on the input-output theory \cite{Walls}, the output (reflected) field of the cavity is obtained as
\begin{equation}\label{INOUT}
\varepsilon_{out}(t)=\kappa^\prime\langle a(t)\rangle-\varepsilon_{in}(t),
\end{equation}
where $\kappa^\prime\ll\kappa$ is the rate at which the cavity is weakly coupled to the drive port. Besides, the input field generated by the coupling laser and probe signal is given by $\varepsilon_{in}(t)=-i\eta-i\zeta^\ast e^{-i\omega_{pc}t}$, and consequently the output (reflected) field of cavity can be obtained from Eqs.(\ref{matL}) and (\ref{INOUT}) as follows
\begin{subequations}
	\begin{align}
\varepsilon_{out}(t)=A_c + A_{aS} e^{-i\omega_{pc}t} + A_{S}  e^{i\omega_{pc}t},\label{epsoutt}\\
A_c=i\eta+\kappa^\prime\alpha,\\
A_{aS}=i\zeta^\ast\Big(1-i\kappa^\prime\tilde G_R^{aa^\dag}(\omega_{pc})\Big),\label{Aas}\\
A_{S}=\kappa^\prime\zeta \tilde G_R^{aa}(-\omega_{pc}),\label{As}
	\end{align}
\end{subequations}
where $A_{c}$ is the amplitude of the carrier (central) band, while $A_{aS}$ and $A_{S}$ are, respectively, the anti-Stokes and Stokes amplitudes of the cavity output field. Obviously,  $A_{c}$,  $A_{aS}$, and  $A_{S}$ are nothing except for the Fourier coefficients of $\varepsilon_{out}(t)$ at frequencies $\omega_c$, $\omega_c+\omega_{pc}$, and $\omega_c-\omega_{pc}$ in the laboratory frame. As is well-known in quantum optomechanics, in the red detuned regime ($\omega_c<\omega_{0}$) the anti-Stokes sideband is enhanced while the Stokes sideband is attenuated considerably \cite{Aspelmeyer}. Since our study is limited to the red detuned regime where the system is stable, the amplitude of anti-Stokes sideband, i.e., Eq.(\ref{Aas}), is important as it is the reflected field amplitude at the probe frequency in the laboratory frame. 

In an OMIT experiment, what is measured experimentally is the power spectrum \cite{Papoulis} of the cavity output field of (Eq.(\ref{epsoutt})) at the probe frequency which is proportional to $|A_{aS}|^2=|\zeta|^2\mathcal{R}(\omega_{pc})$ in the rotating frame, where 
\begin{equation}\label{Rexact}
\mathcal{R}(\omega_{pc})=|1-i\kappa^\prime\tilde G_R^{aa^\dag}(\omega_{pc})|^2
\end{equation}
is the power reflection coefficient of the cavity \cite{twophonon} at probe frequency (in the rotating frame) which represents the response of the system to the input probe signal. In the limit $\kappa^\prime\ll\kappa$, it can be easily shown that $\mathcal{R}$ can be obtained approximately as
\begin{equation}\label{Rapp}
\mathcal{R}(\omega_{pc})\approx1-2\pi\kappa^\prime\mathcal{\rho}(\omega_{pc})
\end{equation}
where, $\rho(\omega_{pc})$ is the density of states \cite{Ryndyk} of the optical mode of the cavity (in the rotating frame) which is defined as
\begin{equation}\label{DoS}
\rho(\omega_{pc})=-\frac{1}{\pi}{\rm{Im}}[\tilde{G}_R^{aa^\dag}(\omega_{pc})].
\end{equation}
The density of states of Eq.(\ref{DoS}) represents the number of normal modes of the optical field of the cavity that are around $\omega_{pc}$ in the rotating frame which corresponds to the number of normal modes around $\omega_p$ in the laboratory frame.

\section{Nonlinear equations of motion of Green's functions}\label{EOM}
As is seen from Eq.(\ref{Mat}), in order to obtain the linear response of the OMS to the input time-dependent perturbation, one needs to have the optical field retarded Green's functions defined by Eqs.(\ref{Gaa}), (\ref{Gaad}). Based on the GLRT all the system Green's functions can be obtained through a system of differential equations, known as Green's functions equations of motion, which are derived from the QLEs \cite{aliGreen}. For this purpose, we first introduce the retarded Green's functions of the system generally as the following compact form
\begin{equation}\label{GijR}
G^{ij}_{R}(t)\equiv-i\theta (t)\langle [u^i(t),u^j(0)]\rangle_0,
\end{equation}
where the instant $t=0$ in Eq.(\ref{GijR}) has been considered as an arbitrary instant of time when the system has reached the steady state. Furthermore, we have assumed that the OMS has been interacting with the environment since $t=-\infty$. As is clearly seen from Eq.(\ref{GijR}), the optical field Green's functions given in Eqs.(\ref{Gaa}), (\ref{Gaad}) are nothing except for $G^{11}_R(t)$ and $G^{12}_R(t)$.

In order to obtain the Green's functions equations of motion, it is enough to take the time derivative of the system Green's functions of Eq.(\ref{GijR}) as follows
\begin{eqnarray}
\dot G^{ij}_{R}(t)=-i\delta(t)\langle[u^i(t),u^j(0)]\rangle_0-i\theta(t)\langle[\dot u^i(t),u^j(0)]\rangle_0,
\end{eqnarray}
where $\delta(t)$ is the Dirac delta function.
Substituting $\dot u^{i}(t)$ with the right hand side of nonlinear QLEs of Eq.(\ref{compEOM}), we obtain
\begin{align}\label{Gt1}
\dot G^{ij}_{R}(t) = & -i\delta(t)\langle[u^i(t),u^j(0)]\rangle_0-i\theta(t)\chi^{i}_{m}\langle[u^m(t),u^j(0)]\rangle_0\nonumber\\
&-i\theta(t)\Gamma^{i}_{mn}\langle[u^m(t)u^n(t),u^j(0)]\rangle_0-i\theta(t)\langle[\xi^i(t),u^j(0)]\rangle_0.\nonumber\\
\end{align}
It should be noted that the quantum noise operator at any instant of time $t$ commutes with the system operators at earlier times \cite{Gardiner}. It means that $[\xi^i(t),u^j(0)]=0$ for $t>0$ and consequently the last term in Eq.(\ref{Gt1}) is zero. Besides, by introducing the following definitions
\begin{subequations}
	\begin{eqnarray}
	\mathcal{J}^{ij}&\equiv&\langle[u^i(0),u^{j}(0)]\rangle_0,\label{Jdef}\\
	\mathcal{P}^{mnj}(t)&\equiv&-i\theta(t)\langle[u^m(t)u^n(t),u^j(0)]\rangle_0,\label{Pdef}
	\end{eqnarray}
\end{subequations}
and considering the definition of the Green's functions equations of motion, Eq.(\ref{Gt1}) can be read as
\begin{eqnarray}\label{Gt2}
\dot G^{ij}_{R}(t)=-i\delta(t)\mathcal{J}^{ij}+\chi^{i}_{m}G^{mj}_{R}(t)+\Gamma^{i}_{mn}\mathcal{P}^{mnj}(t).
\end{eqnarray}
As is seen from Eq.(\ref{Gt2}), the Green's function equations of motion depend on the new functions of time, i.e., $\mathcal{P}^{mnj}(t)$. Therefore, it is necessary to obtain their equations of motions by taking the time derivative as follows
\begin{eqnarray}\label{Pt1}
\dot{\mathcal{P}}^{mnj}(t)=-i\delta(t)\langle[u^m(t)u^n(t),u^j(0)]\rangle_0\nonumber\\
-i\theta(t)\langle[\dot u^m(t)u^n(t),u^j(0)]\rangle_0\nonumber\\
-i\theta(t)\langle[u^m(t)\dot u^n(t),u^j(0)]\rangle_0.
\end{eqnarray}
The presence of delta function in the first term of Eq.(\ref{Pt1}) means that it is zero for $t\neq 0$. However, it can be easily shown that the first term is also zero for $t=0$ because expansion of the commutator contains terms like $\langle u^m(0)[u^n(0),u^j(0)]\rangle$ in which the inside commutator can be brought out of the mean value and $\langle u^m(0)\rangle$ is zero. In this way, the first term is always zero. On the other hand, by substituting the time derivative of the components of $u$ with the right hand side of Eq.(\ref{compEOM}) in the second and third terms, we will have
\begin{eqnarray}\label{Pt2}
\dot{\mathcal{P}}^{mnj}(t)&=&-i\theta(t)\chi^m_l\langle[u^l(t)u^n(t),u^j(0)]\rangle_0\nonumber\\
&&-i\theta(t)\chi^n_l\langle[u^m(t)u^l(t),u^j(0)]\rangle_0\nonumber\\
&&-i\theta(t)\Gamma^m_{lk}\langle[u^l(t)u^k(t)u^n(t),u^j(0)]\rangle_0\nonumber\\
&&-i\theta(t)\Gamma^n_{lk}\langle[u^m(t)u^l(t)u^k(t),u^j(0)]\rangle_0\nonumber\\
&&-i\theta(t)\langle[\xi^m(t)u^n(t),u^j(0)]\rangle_0\nonumber\\
&&-i\theta(t)\langle[u^m(t)\xi^n(t),u^j(0)]\rangle_0.
\end{eqnarray}
Based on the definition of Eq.(\ref{Pdef}), the first two terms in the right hand side of Eq.(\ref{Pt2}) can be written in terms of $\mathcal{P}$ functions. The third and fourth terms which contain 4-fold correlations, lead to an infinite hierarchy of equations. In order to truncate the hierarchy we should make an approximation at this stage. Based on the Appendix \ref{apA}, by neglecting the correction term $\mathcal{C}^{(4)}$, the expectation values in the third term of Eq.(\ref{Pt2}) can be approximated as
\begin{eqnarray}
\langle[u^l(t)u^k(t)u^n(t),u^j(0)]\rangle_0 &
\approx \langle u^l(t)u^k(t)\rangle_0\langle [u^n(t),u^j(0)]\rangle_0 \nonumber\\ 
&+\langle u^l(t)u^n(t)\rangle_0\langle [u^k(t),u^j(0)]\rangle_0\nonumber\\
&+\langle u^k(t)u^n(t)\rangle_0\langle[u^l(t),u^j(0)]\rangle_0.\nonumber  
\end{eqnarray}
A similar expression like the right hand side of the above equation can be obtained for the expectation values in the fourth term of Eq.(\ref{Pt2}).

Finally, it can be shown that the last two terms in Eq.(\ref{Pt2}) containing quantum noises are zero. For this purpose, first consider the expectation value in the fifth term which can be expanded as follows
\begin{eqnarray}
\langle[\xi^m(t)u^n(t), u^j(0)]\rangle_0 =\langle[\xi^m(t), u^j(0)]u^n(t)\rangle_0  \nonumber\\
+ \langle\xi^m(t)[u^n(t), u^j(0)]\rangle_0, \nonumber
\end{eqnarray}
where the first term is zero because the quantum noise operator at any time $t$ commutes with the system operators at the earlier times as was mentioned before. The last term is also zero because based on the Appendix \ref{apA} it can be approximated as follows
\begin{eqnarray}
\langle\xi^m(t)[u^n(t), u^j(0)]\rangle_0 \approx \langle\xi^m(t)\rangle_0\langle [u^n(t), u^j(0)]\rangle_0 = 0. \nonumber
\end{eqnarray}

In this way, the equation of motion of $\mathcal{P}$ functions, i.e., Eq.(\ref{Pt2}), can be read as follows
\begin{align}\label{Pt3}
\dot{\mathcal{P}}^{mnj}(t) & =\chi^m_l\mathcal{P}^{lnj}(t) + \chi^n_l\mathcal{P}^{mlj}(t)\nonumber\\
& +\Gamma^m_{lk}\langle u^l(t)u^k(t)\rangle_0 G^{nj}_R(t) + \Gamma^m_{lk}\langle u^l(t)u^n(t)\rangle_0 G^{kj}_R(t)\nonumber\\
& +\Gamma^m_{lk}\langle u^k(t)u^n(t)\rangle_0 G^{lj}_R(t) + \Gamma^n_{lk}\langle u^l(t)u^k(t)\rangle_0 G^{mj}_R(t)\nonumber\\
& +\Gamma^n_{lk}\langle u^m(t)u^k(t)\rangle_0 G^{lj}_R(t) + \Gamma^n_{lk}\langle u^m(t)u^l(t)\rangle_0 G^{kj}_R(t),\nonumber\\
\end{align}
where the two-fold expectation values, can be written in terms of the symmetric tensor  $V^{ij}(t) = \frac{1}{2}\langle u^i(t)u^j(t)+u^j(t)u^i(t)\rangle_0$ in  the following form
\begin{equation}\label{mu-VJ}
\langle u^i(t)u^j(t)\rangle_0 = V^{ij}(t) + \mathcal{J}^{ij}.
\end{equation}
The appearance of $\mathcal{J}$ in Eq.(\ref{mu-VJ}) is a consequence of the fact that the field commutator at equal times are the same at all times. Since the two-fold expectation values in Eq.(\ref{Pt3}) should be calculated in the steady state, i.e., $t=0$, the components $V^{ii}(t)$ should be calculated at steady state which are obtained from the Lyapunov equation 
\begin{eqnarray}\label{Lyap}
\chi V + V\chi^{T} = -\mathcal{D},
\end{eqnarray}
where 
\begin{equation}
\mathcal{D}=\left(\begin{array}{cccc}
0 & \kappa/2 & 0 & 0 \\
\kappa/2 & 0 & 0 & 0 \\
0 & 0 & 0 & (1+2n_{th})\gamma/2\\
0 & 0 & (1+2n_{th})\gamma/2 & 0\\
\end{array}\right),
\end{equation}
is the diffusion matrix with $n_{th}$ being the mean number of thermal phonons of the mechanical reservoir. In this way, all the coefficients in Eq.(\ref{Pt3}) are determined.

Therefore, Eqs.(\ref{Pt3}) and (\ref{Gt2}) form a system of linear ordinary differential equations with constant coefficients, from  which all the retarded Green's functions of the system can be derived. By taking their Fourier transformation, they are transformed into a set of algebraic equations as follows
\begin{align}\label{GPw}
&i\omega \tilde{G}^{ij}_{R}(\omega) +\chi^{i}_{m}\tilde{G}^{mj}_{R}(\omega)+\Gamma^{i}_{mn}\mathcal{\tilde{P}}^{mnj}(\omega)=i\mathcal{J}^{ij},\\
&i\omega\mathcal{\tilde{P}}^{mnj}(\omega)+\chi^m_l\mathcal{\tilde{P}}^{lnj}(\omega) + \chi^n_l\mathcal{\tilde{P}}^{mlj}(\omega)\nonumber\\
&+ \Gamma^m_{lk}(V^{lk}+\mathcal{J}^{lk})\tilde{G}^{nj}_R(\omega) + \Gamma^m_{lk}(V^{ln}+\mathcal{J}^{ln})\tilde{G}^{kj}_R(\omega)\nonumber\\
&+\Gamma^m_{lk}(V^{kn}+\mathcal{J}^{kn})\tilde{G}^{lj}_R(\omega) + \Gamma^n_{lk}(V^{lk}+\mathcal{J}^{lk})\tilde{G}^{mj}_R(\omega)\nonumber\\
&+\Gamma^n_{lk}(V^{mk}+\mathcal{J}^{mk})\tilde{G}^{lj}_R(\omega) + \Gamma^n_{lk}(V^{ml}+\mathcal{J}^{ml})\tilde{G}^{kj}_R(\omega)=0.\nonumber
\end{align}
The set of Eqs.(\ref{GPw}) is a system of 80 non-homogeneous linear algebraic equations which can be solved numerically. However, since we are interested in the optical response of the system we do not need to solve all the 80 equations simultaneously. For example for obtaining $\tilde{G}_{R}^{aa^{\dagger}}(\omega)$ it is enough to set $j=2$ and consequently solve just 20 out of 80 equations.

\section{results and discussion}\label{RD}
In order to see under which conditions the nonlinear interaction, i.e., the last term of Eq.(\ref{HS}), affects the linear response and density of states of the system, it should be noted that the OMS in the absence of nonlinear interaction is a dynamical system consisting of two coupled oscillators (optical and mechanical modes indicated by $a$ and $b$ operators) interacting with each other through a radiation pressure interaction. As is seen from Eqs.(\ref{compEOM}) and (\ref{chi}), the linear part of equations of motion are coupled to each other. However, there are  normal variable, named $\delta C_+$ and $\delta C_-$, which lead to a system of decoupled equations in the absence of nonlinear interaction (see Eqs. (9) and (10) of \cite{Signature}). As has been shown in \cite{Parametric-Down-Conversion}, the normal variables are related to original variables $(\delta a,\delta b)$ through a canonical transformation and the system Hamiltonian can be written in terms of the normal variables as 
\begin{widetext}
	
	\begin{figure}
		\centering
		\includegraphics[width=6.5cm]{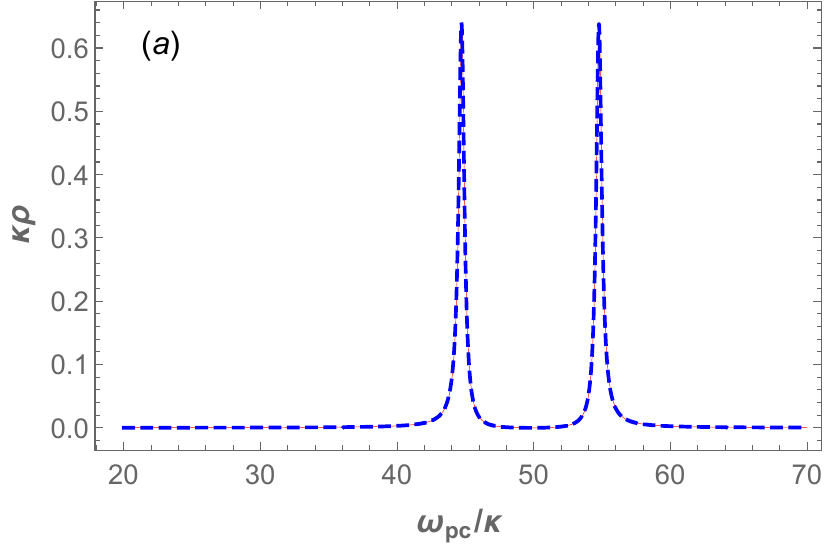}
		\includegraphics[width=6.5cm]{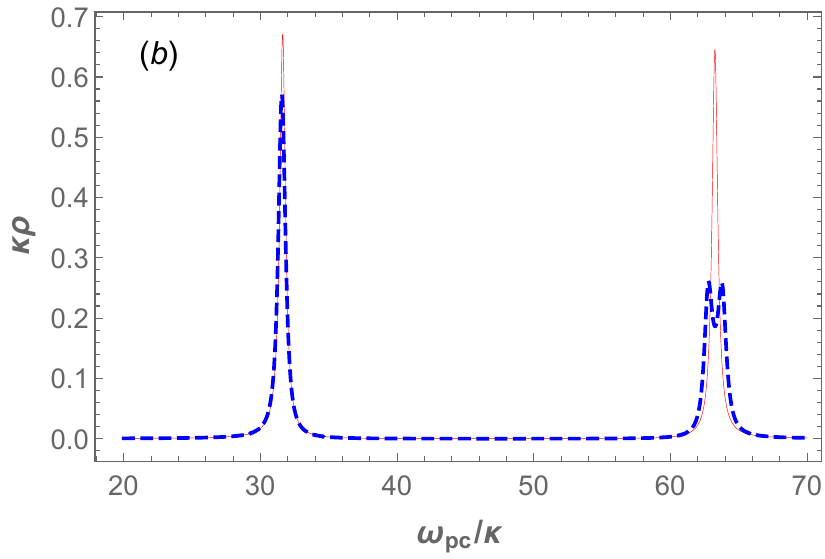}
		\includegraphics[width=6.5cm]{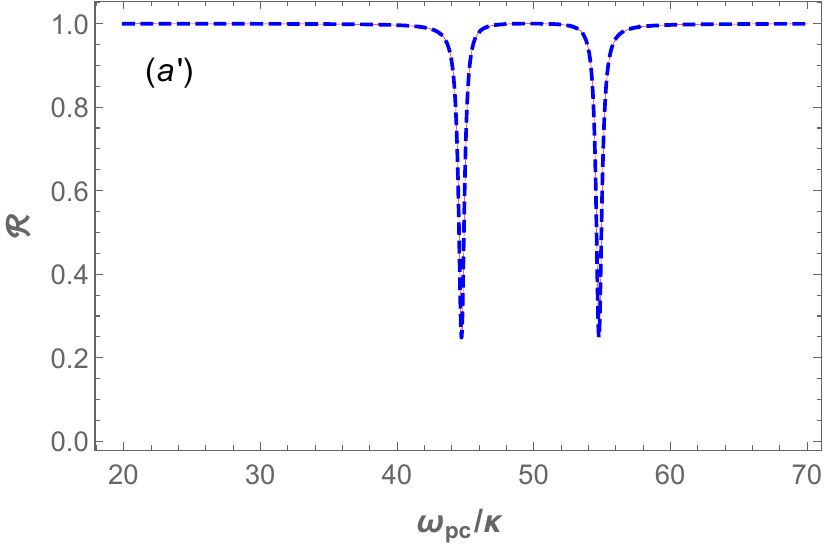}
		\includegraphics[width=6.5cm]{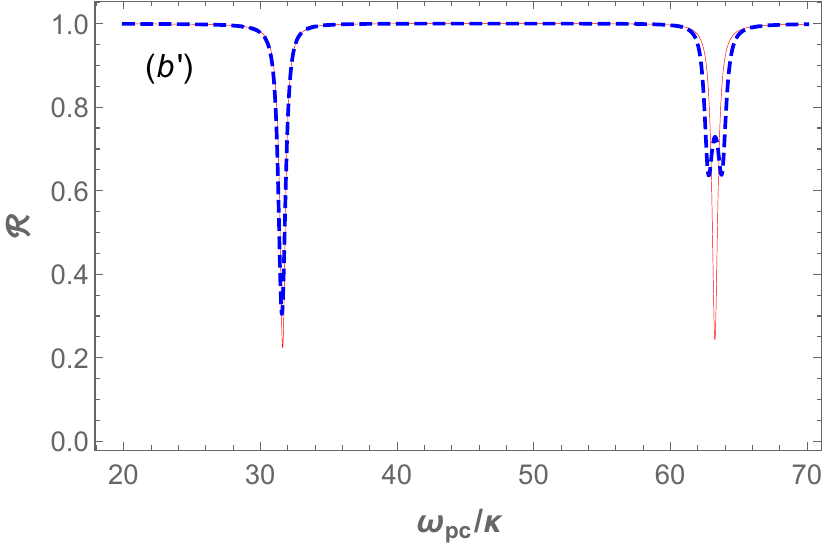}
		\caption{
			(Color online) The normalized density of states $\kappa\rho(\omega_{pc}/\kappa)$ as well as the cavity power reflection coefficient $\mathcal{R}(\omega_{pc}/\kappa)$ when the system is in the red detuned regime of $\Delta=\omega_m$. The system parameters have been chosen as $\omega_m=50\kappa$,  $g_0=1\kappa$, $\gamma=10^{-4}\kappa$, $\kappa^\prime=0.25\kappa$, and $n_{th}=1$. (a,a$'$) The enhanced optomechanical coupling is fixed at $g=5\kappa$. (b,b$'$) The enhanced optomechanical coupling is fixed at $g=15\kappa$. The red solid and the blue dashed curves correspond, respectively, to the absence and presence of the nonlinearity.
		}
		\label{fig1}
	\end{figure}
	
\end{widetext}

\begin{subequations}
	\begin{align}
		H_S&=H_L + H_{NL},\\
		H_L&=\hbar\omega_{-}\delta C^\dagger_{-}\delta C_{-}+\hbar\omega_{+}\delta C^\dagger_{+}\delta C_{+},\label{HL}\\
		H_{NL}&=\hbar g'  \delta C^{\dagger2}_{-}\delta C_{+} +H.c+....\label{HNL},
	\end{align}
\end{subequations}
where $g'$ is a coupling parameter proportional to $g$ (see equation (3) of \cite{Parametric-Down-Conversion}), and $\omega_{\pm}$ are normal frequencies of the system. In order to obtain the normal frequencies of the system one should solve for the eigenvalue problem of the matrix $\chi$ of Eq.(\ref{chi}) in which $\kappa$ and $\gamma$ has been set to zero. In this way the normal frequencies are obtained as follows
\begin{eqnarray}\label{wpm}
	\omega_{\pm}=\frac{1}{\sqrt{2}}\Big(\Delta^2+\omega_m^2\pm\sqrt{(\Delta^2-\omega_m^2)^2+16g^2\omega_m\Delta}\Big)^{1/2}.\label{wI}
\end{eqnarray}
As is seen from Eq.(\ref{HNL}), the nonlinear part remains nondiagonal with many terms where just those have been shown in Eq.(\ref{HNL}) that have a dominant effect on the linear response of the system. The important point is that the two shown terms with the coupling parameter $g'$ in Eq.(\ref{HNL}) become time independent in the interaction picture if the condition $\omega_{+}=2\omega_{-}$ is satisfied which makes them be the dominant nonlinear interaction in the rotating wave approximation (RWA), while other terms become time dependent and have negligible contribution in the RWA.

Before presenting our results in the rest of this section, it should be emphasized that our approach in the present work has not been based on the normal variable concept at all; but we have made used of it just as an ancillary tool for the physical interpretation of our results obtained in the continuation of this section. In fact, one of the most important advantages of our approach over those based on normal variables \cite {Signature,Parametric-Down-Conversion,Lemonde2013} is that in the GLRT all the calculations are done in terms of the original variables $(\delta a,\delta b)$ without any necessity of a canonical transformation to the normal variables.

In order to see the effects of nonlinear optomechanical interaction on the system linear response, we solve Eqs.(\ref{GPw}) together with Eq.(\ref{Lyap}) numerically by fixing $j=2$. It should be emphasized that by solving the Green's functions equations of motion, i.e., Eqs.(\ref{GPw}), the contributions of all nonlinear terms of $H_{NL}$ (even those that have not been shown in Eq.(\ref{HNL})) are taken into account. 

For this purpose, in Fig.\ref{fig1} we have plotted the normalized density of states $\kappa\rho$ as well as the cavity power reflection coefficient $\mathcal{R}$ versus the normalized frequency $\omega_{pc}/\kappa$ while the system is in the red detuned regime of $\Delta=\omega_{m}$. It has been assumed that the single-photon optomechanical coupling is $g_0=1\kappa$, and the frequency of the mechanical mode is $\omega_m=50\kappa$ so that the OMS lies in the resolve sideband regime ($\omega_m>\kappa$) with the mechanical damping rate being $\gamma=10^{-4}\kappa$. It should be noted that for $\Delta=\omega_m$, the resonance condition $\omega_{+}=2\omega_{-}$ leads to a specified value of $g=0.3\omega_m$ based on Eq.(\ref{wpm}). In Figs.\ref{fig1}(a) and \ref{fig1}(a$'$), the response of the system has been demonstrated for $g=5\kappa$, where the condition $\omega_{+}=2\omega_{-}$ is not satisfied, while in Figs.\ref{fig1}(b) and \ref{fig1}(b$'$) the enhanced optomechanical coupling has been fixed at $g=15\kappa=0.3\omega_m$, where the condition $\omega_{+}=2\omega_{-}$ is satisfied. The red solid curves show the response of the system in the absence of $H_{NL}$ while the blue dashed curves show the system response in the presence of the nonlinearity.

It should be noted that in Fig.\ref{fig1} as well as the following ones we have demonstrated the system behavior by the red-solid curves based on an approximate model in which the nonlinear optomechanical interaction (the last term in Eq.(\ref{HS}) has been eliminated from the system Hamiltonian. On the contrary, we have demonstrated the system behavior by the blue-dashed curves based on an exact model in which the nonlinear optomechanical interaction (the last term in Eq.(\ref{HS}) has been considered in the system Hamiltonian. In this way, we would like to show the difference between the predictions based on the exact model (indicated by blue-dashed curves) and those of the approximate model (indicated by red-solid curves).

The most important result that is clearly observed from Fig.\ref{fig1} is the fact that unless the resonance condition $\omega_{+}=2\omega_{-}$ is satisfied the effect of nonlinear interaction $H_{NL}$ is not manifested in the system response even when the single-photon optomechanical coupling is of the order of the cavity damping rate $(g_0=1\kappa)$. In other words, in Figs.\ref{fig1}(a) and \ref{fig1}(a$'$), where the system is out of resonance condition the linear response of the system in the presence of the nonlinearity (blue dashed curves) coincides with that in the absence of it (red solid curves). Instead, in Figs.\ref{fig1}(b) and \ref{fig1}(b$'$), where the resonance condition is satisfied the presence of the nonlinear interaction leads to a hybridization in the second peak (dip) of the system density of states (reflection coefficient).

\begin{figure}
		\centering
		\includegraphics[width=6.5cm]{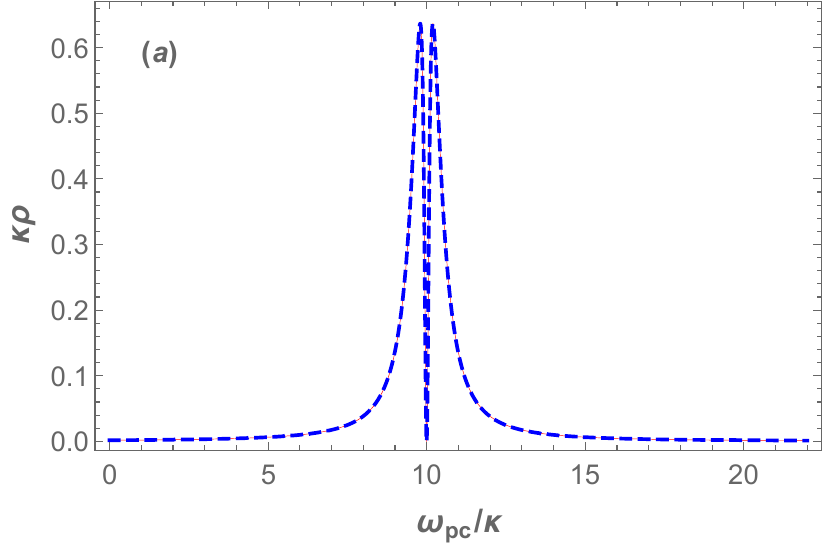}
		\includegraphics[width=6.5cm]{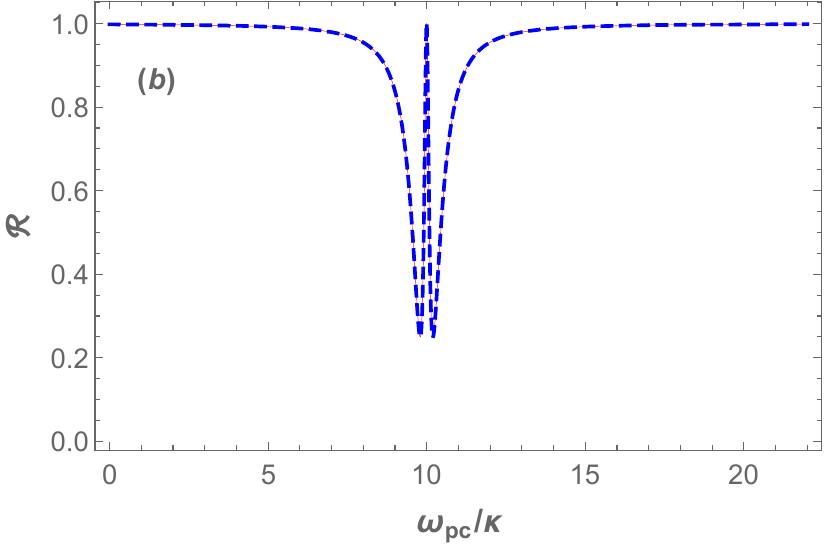}
		\caption{
			(Color online) (a) The normalized density of states $\kappa\rho(\omega_{pc}/\kappa)$, and (b) the cavity power reflection coefficient $\mathcal{R}(\omega_{pc}/\kappa)$ for an OMS with $\omega_{m}=10\kappa$ and $g_0=0.2\kappa$, $\gamma=10^{-4}\kappa$, $\kappa^\prime=0.25\kappa$, and $n_{th}=1$. The system is in the red detuned regime of $\Delta=\omega_m$ with $g=0.2\kappa$. The red solid and the blue dashed curves correspond, respectively, to the absence and presence of the nonlinearity.
		}
		\label{fig2}
\end{figure}

The other point that should be noted is that since in both cases demonstrated in Fig.\ref{fig1} the cavity is strongly driven by the coupling laser $(g>\kappa)$, the system lies in the normal mode splitting (NMS) regime \cite{Dobr}, where the two normal modes of the system are completely separated and a wide transparency window is manifested between them.  As is seen from the red solid curves in Fig.\ref{fig1}, the density of states peaks (reflection coefficient dips) appear at $\omega_{pc}=\omega_{\pm}$ which occur at $\omega_{\pm}=\omega_m\sqrt{1\pm2g/\omega_m}$ for $\Delta=\omega_m$ in the absence of the nonlinear interaction. It is a completely expected result because the OMS is a two-mode dynamical system with two normal frequencies $\omega_{\pm}$. That is why in  Figs.\ref{fig1}(a) and \ref{fig1}(a$'$) where $g=5\kappa$, the normal frequencies appear at $\omega_-\approx 44.7\kappa$ and $\omega_+\approx 54.8\kappa$. In fact, when the probe-coupling detuning ($\omega_{pc}$) is on resonance with one of normal mode frequencies $(\omega_{\pm})$, where the system density of states maximizes (Figs.\ref{fig1}(a) and \ref{fig1}(b)), the OMS absorbs the maximum energy from the input source, and consequently the output reflected field $(\mathcal{R})$ becomes minimum (Figs.\ref{fig1}(a$'$)) and \ref{fig1}(b$'$)). Otherwise, since the density of states is zero, no energy is absorbed by OMS and consequently the reflection maximizes. 

On the contrary, to illustrate the situation in the weakly driving regime, in Fig.\ref{fig2} we have demonstrated the linear response of an OMS with $\omega_{m}=10\kappa$ in the red detuned regime of $\Delta=\omega_{m}$ for $g=0.2\kappa$ while the single-photon optomechanical coupling is $g_0=0.2\kappa$. In this case, since $g<\kappa/4$ the system lies in the OMIT regime \cite{Fano What} so that a narrow transparency window appears at the anti-resonance frequency $\omega_{pc}=\omega_m=10\kappa$, where the optical density of states is zero, and consequently the output reflected field maximizes because the OMS absorbs no energy from the input source since the system has no normal mode at that frequency to absorb energy. It can also be seen from Eq.(\ref{Rapp}), which shows that $\mathcal{R}=1$ if $\rho=0$. 

\begin{figure}
	\centering
	\includegraphics[width=6.5cm]{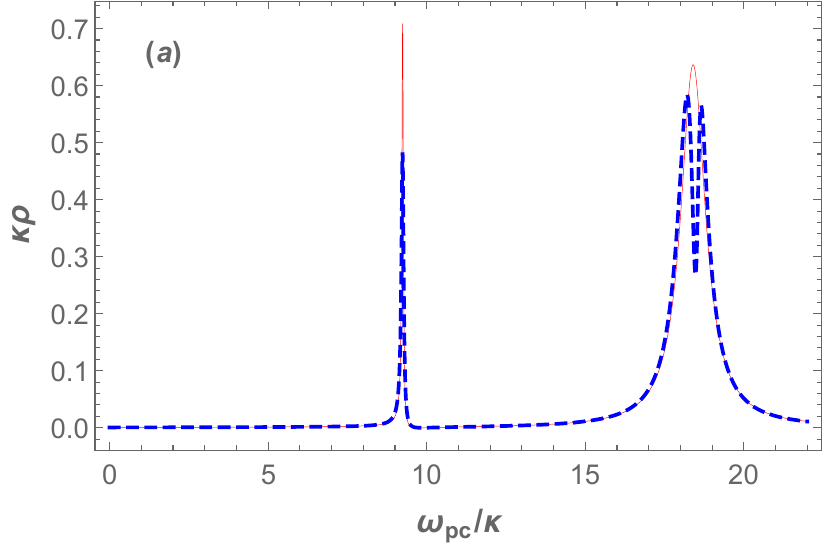}
	\includegraphics[width=6.5cm]{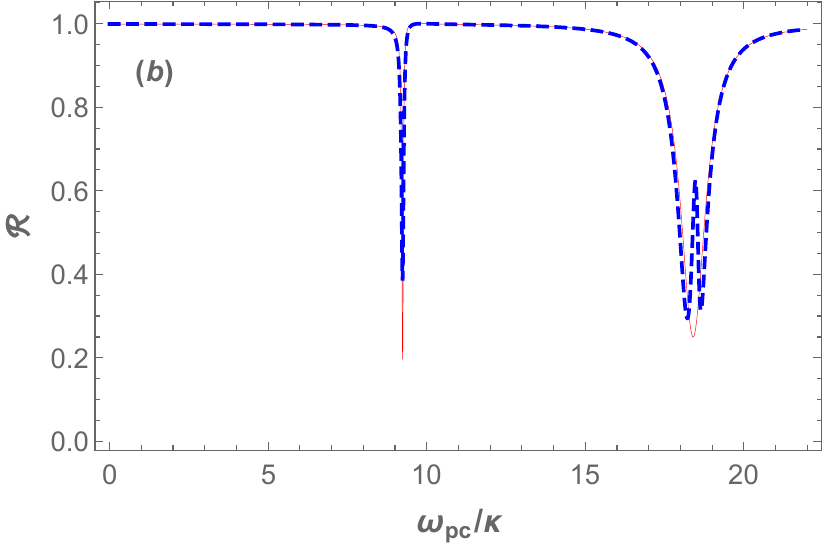}
	\caption{
		(Color online) (a) The normalized density of states $\kappa\rho(\omega_{pc}/\kappa)$, and (b) the cavity power reflection coefficient $\mathcal{R}(\omega_{pc}/\kappa)$ for an OMS with $\omega_{m}=10$ and $g_0=0.5\kappa$, $\gamma=10^{-4}\kappa$, $\kappa^\prime=0.25\kappa$, and $n_{th}=1$. The system is in the red detuned regime of $\Delta=1.8\omega_m$ with $g=2.2\kappa$. The red solid and the blue dashed curves correspond, respectively, to the absence and presence of the nonlinearity.
	}
	\label{fig3}
\end{figure}

Besides, since $g=0.2\kappa$ does not satisfy the resonance condition $\omega_{+}=2\omega_{-}$ the effect of the nonlinearity is not manifested although the OMS is in the single-photon regime, where $\alpha=g/g_0=1$. That is why the blue dashed curve (corresponding to the presence of nonlinearity) coincides with the red solid line (corresponding to the absence of nonlinearity). Interestingly, this result confirms the validity of the linearization approximation which is often used in the weak coupling regime where some important phenomena like OMIT occurs. It means that the linearization approximation is valid for the description of OMIT even in the single photon regime. Generally, the linearization approximation is valid as far as $\Delta$ and $g$ take the values that do not satisfy the condition $\omega_+=2\omega_-$.

The anti-resonance is an interesting phenomenon which occurs in a dynamical system consisting of coupled oscillators that is driven by an external source \cite{Belbasi,Joe}. In a two-mode driven dissipative system consisting of two coupled oscillators (like a bare OMS), there is an anti-resonance frequency at which the oscillation amplitude of the oscillator that is directly driven by the external source (the cavity optical mode in OMS) becomes zero when the frequency of the external source coincides with the natural frequency of the second oscillator $(\omega_m)$. It occurs when the damping rate of the second oscillator is negligible in comparison to that of the first one which is directly driven by the external source $(\gamma\ll\kappa)$.  Under this condition, when the frequency of the external source gets near to the anti-resonance point from lower values, the phase of the first oscillator suffers a sudden change and becomes out of phase with the second one so that the motion of the first oscillator is quenched effectively by the second one. 

So far, we have studied the situation of $\Delta=\omega_m$, where the effective cavity detuning is equal to the mechanical frequency. As has been already mentioned, in this situation the system exhibits two symmetric normal modes with equal linewidths while an anti-resonance point appears in the middle of them in the absence of nonlinear interaction, which is the origin of the OMIT phenomenon in the weak-driving regime. Now, we would like to study the situation of $\Delta\neq\omega_m$ corresponding to the Fano resonance, which was first discovered by Ugo Fano \cite{Fano rev} in the absorption spectrum of Rydberg atoms exhibiting sharp asymmetric profiles.

For this purpose, in Fig.\ref{fig3} we have demonstrated the linear response of an OMS with $\omega_{m}=10\kappa$ and $g_0=0.5\kappa$ while the system is in the red detuned regime of $\Delta=1.8\omega_m$ for which the resonance condition $(\omega_+=2\omega_-)$ is satisfied while the enhanced optomechanical coupling is fixed at $g=2.2\kappa$. The red solid and the blue dashed curves correspond, respectively, to the absence and presence of the nonlinearity. The important point in the phenomenon of Fano resonance is that the anti-resonance which occurs again at $\omega_{pc}=\omega_m$, gets nearer to one of the normal modes frequency which makes the profile of that normal mode become narrow and asymmetric. That is why, the first peak (dip) in Fig.\ref{fig2} corresponding to the first normal mode at frequency $\omega_-\approx 9.2\kappa$, becomes narrow and asymmetric because it is near the anti-resonance frequency $(\omega_m=10\kappa)$. Interestingly, because the resonance condition $\omega_+=2\omega_-$ is satisfied, the second normal mode peak hybridizes in the presence of the nonlinearity (the blue dashed curves) even for the single-photon coupling of $g_0=0.5\kappa$.

\begin{figure}
	\centering
	\includegraphics[width=8cm]{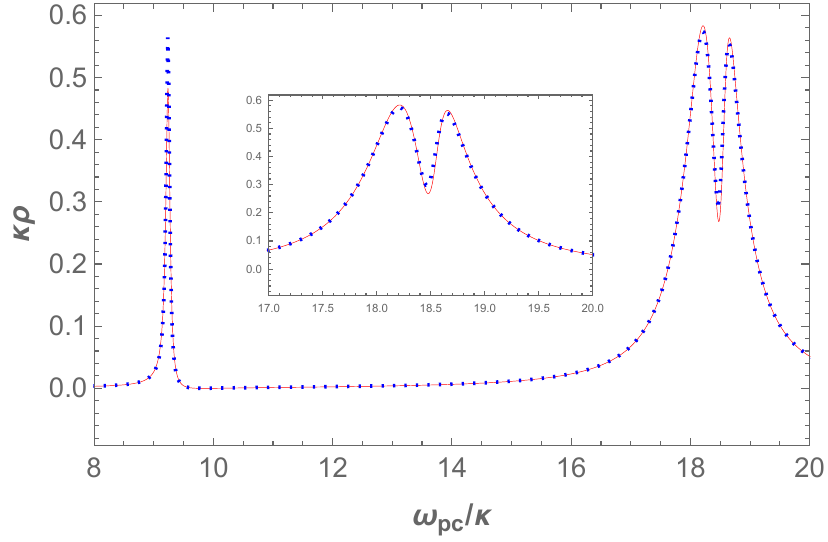}
	\caption{
		(Color online) Comparison between the analytical method of GLRT (the red solid curve) and the numerical calculation of master equation (the blue dotted curve) for an OMS with characteristics presented in Fig.\ref{fig3}. The inset shows a maginification of the second hybridized peak.
	}
	\label{fig4}
\end{figure}

It should be noted that any additional nonlinear interaction in the system Hamiltonian can change the normal modes of the linear system. As has been shown in Ref.\cite{singlephoton} in the weak-drive and very strong coupling regime $(g_0\sim\omega_m)$, the nonlinear interaction leads to the increase in the number of peaks. However, in the present work, where we have studied the OMS in the strong-drive and weak coupling regime, the nonlinear interaction is weak and we can consider it as a perturbation in our calculations. It should be noted that a weak perturbation does not add any new normal mode to the system (new peaks to density of states) but it only leads to some slight modifications in the previous normal modes of the system. That is why only the peaks of the density of states are modified in our results. For example, the height of first peak is reduced and the second peak is hybridized (is split) while other parts of the density of states curve remains unchanged.

Furthermore, there are other references \cite{Signature,Parametric-Down-Conversion} which have also studied the OMS in strong-drive and weak coupling regime. Although those studies have been done in the same regime investigated by us, but the main difference between our work and theirs is that they have investigated the OMS in terms of normal variable and polariton modes to show under which conditions the polariton modes have photon-like, phonon-like, or mixed photon-phonon behaviors. On the contrary, our calculations have been based on the original modes $(a,b)$ and the concept of polaritons have just introduced as an ancillary tool for a physical interpretation of our results as was mentioned before. Our aim has been to show very straightforwardly through the GLRT how an OMS can respond to a weak perturbation and predict the output cavity field behavior in an experimental setup.

Finally, in order to show the reliability of our analytical approach based on GLRT, in Fig.\ref{fig4} we have compared the density of states of an OMS with parameters given in Fig.\ref{fig3} obtained from the GLRT method (the red solid curve) with that obtained by the numerical calculation of master equation (the blue dotted curve) using QuTiP \cite{qutip1,qutip2}.The master equation that we have solved numerically to compare our results with is given by 
\begin{align}\label{master}
\frac{\partial}{\partial t}\rho(t)&=-\frac{i}{\hbar}[H_s,\rho]+\frac{\kappa}{2}(2\delta a\rho \delta a^{\dagger}-\delta a^{\dagger}\delta a\rho-\rho \delta a^{\dagger}\delta a)\nonumber\\
&+\frac{\gamma}{2}(n_{th}+1)(2\delta b\rho \delta b^{\dagger}- \delta b^{\dagger}\delta b\rho-\rho \delta b^{\dagger}\delta b)\nonumber\\
&+\frac{\gamma}{2}n_{th}(2\delta b^{\dagger}\rho \delta b-\delta b \delta b^{\dagger}\rho-\rho \delta b\delta b^{\dagger}),
\end{align}
where $H_s$ is the system Hamiltonian consisting of the nonlinear interaction given by Eq.(\ref{HS}). It should be reminded that in Eq.(\ref{master}) the number of thermal photons has been assumed to be zero ($n_{ph}=0$).  As is seen clearly, there is a very good coincidence between the two methods. In Appendix \ref{apB} we have explained how the form of density of states given by Eq.(\ref{DoS}) should be modified to be calculable by the QuTiP.

\section{conclusion}\label{cln}
In this article, we studied the linear response of a driven-dissipative OMS to a weak probe laser in the nonlinear regime where the intrinsic nonlinear optomechanical interaction is no longer negligible. Using the GLRT, we obtained the Stokes and anti-Stokes sidebands of cavity output field as well as the power reflection coefficient and the optical density of states in terms of optical retarded Green's functions. We showed that the nonlinear QLEs of the system lead to an infinite hierarchy in the retarded Green's functions  equations of motion which can be truncated by ignoring 4-fold correlations, while considering the 3-fold correlations as new variables. In this way, the nonlinear equations of motion are transformed to a system of linear equations but with a dimension higher than that of QLEs.

Based on our results, the intrinsic nonlinear effect of optomechanical interaction manifests itself as a hybridization of the second peak of the optical density of states only if the system satisfies a resonance condition in which the frequency of the upper normal mode of the system is twice that of the lower one while the single-photon optomechanical coupling is of the order of cavity linewidth. In this way, the validity of the linearization approximation in quantum optomechanics is confirmed at the off-resonance regime where some important phenomena like OMIT occur. That is why the linearization approximation is valid for the description of OMIT even in the single photon regime. 

In the end, we demonstrated the reliability of our analytical approach through a comparison between the result obtained by the GLRT with that obtained by the numerical calculation of master equation using QuTiP.

\appendix
\section{Truncation of the hierarchy}\label{apA}
In this appendix we would like to indicate how the hierarchy appearing in the Green's functions equations of motion can be truncated. It should be noted that in any nonlinear dynamical system, an n-fold correlation function of the system variables can be decomposed in terms of two-fold correlations (just like the Wick's theorem) plus a correction term arising from higher order correlations due to the existence of the nonlinear interaction. 

For example, a 3-fold correlation can be decomposed as follows
\begin{equation}\label{3fc}
\langle u^iu^ju^k\rangle_0 = \langle u^i\rangle_0\langle u^ju^k \rangle_0 + \langle u^k \rangle_0\langle u^iu^j\rangle_0 +  \langle u^j \rangle_0\langle u^iu^k\rangle_0 + \mathcal{C}^{(3)}\nonumber\\
\end{equation}
where $\mathcal{C}^{(3)} = \mathcal{C}^{(3)}(u^i,u^j,u^k)$ is the correction term and since all the single-fold expectation values, i.e, $\langle u^i \rangle_0 $'s, are zero, Eq.(\ref{3fc}) reduces to
\begin{equation}
\langle u^iu^ju^k\rangle_0 = \mathcal{C}^{(3)}(u^i,u^j,u^k)
\end{equation}
It should be reminded that the $\mathcal{P}$ functions introduced in Eq.(\ref{Pdef}) are nothing except for some $\mathcal{C}^{(3)}$ functions whose effect should be taken into account in our calculations. On the other hand in the case of 4-fold correlations the decomposition can be done as follows
\begin{align}
\langle u^iu^ju^ku^l\rangle_0 &= \langle u^iu^j\rangle_0\langle u^ku^l \rangle_0 +  \langle u^iu^k\rangle_0\langle u^ju^l \rangle_0 \nonumber\\
&+\langle u^iu^l\rangle_0\langle u^ju^k \rangle_0+ \mathcal{C}^{(4)}\nonumber\\
\end{align}
where $\mathcal{C}^{(4)} = \mathcal{C}^{(4)}(u^i,u^j,u^k,u^l)$ is the correction term that should be neglected in order to truncate the hierarchy.

\section{Tip for numerical calculation of density of states using QuTiP}\label{apB}
In order to show how the density of states of Eq.(\ref{DoS}) can be calculated numerically by QuTiP, it should be noted that the Fourier transform of the retarded Green's function of Eq.(\ref{Gaad}) can be written as
\begin{align}
	\tilde{G}_R^{aa^\dag}(\omega)	&= -i \mathcal{I}(\omega)
\end{align}
where $\mathcal{I}(\omega)$ has been defined as
\begin{equation}
	\mathcal{I}(\omega) \equiv \int_{0}^{\infty}d\tau e^{i\omega\tau}\langle [\delta a(\tau+t'),\delta a^\dag(t')]\rangle_0.
\end{equation}
As is seen clearly the density of states can be written as 
\begin{equation}
	\rho(\omega) = \frac{1}{\pi} \rm Re[\mathcal{I}(\omega)].
\end{equation}

What is numerically calculated in QuTiP is the two-time correlation functions 
$\langle \delta a(t+\tau) \delta a^{\dagger}(t)\rangle$
and $\langle  \delta a^{\dagger}(t) \delta a(t+\tau) \rangle$ which are computed by the function \texttt{qutip.correlation\_2op\_1t}. In order to obtain $\mathcal{I}(\omega)$ it is enough to subtract the latter from the former and take its Fourier transform using the function \texttt{qutip.spectrum\_correlation\_fft}. The obtained result is equivalent to the following expression
\begin{equation}
	\int_{-\infty}^{\infty}d\tau e^{i\omega\tau}\langle [\delta a(\tau+t'),\delta a^\dag(t')]\rangle_0,
\end{equation}
which can be easily shown that is just $2\rm Re[\mathcal{I}(\omega)]$. In this way, the density of states which can be considered as the following form
\begin{align}
	\rho(\omega) =\frac{1}{2\pi} \int_{-\infty}^{\infty}d\tau e^{i\omega\tau}\langle [\delta a(\tau+t'),\delta a^\dag(t')]\rangle_0,
\end{align}
can be evaluated numerically by the two above-mentioned functions defined in QuTiP.

\end{document}